\documentclass{CSMA2013}

\title{Modlisation micromagntique du comportement magnto-mcanique}
\author{Frederick Sorel MBALLA MBALLA $^1$, Olivier HUBERT $^1$, Song He $^2$, Sophie DEPEYRE $^2$, Philip MEILLAND $^3$}
\address{%
$^1$ LMT Cachan, ENS Cachan/CNRS/UPMC/PRES UniverSud Paris, mballa@lmt.ens-cachan.fr, hubert@lmt.ens-cachan.fr\\
$^2$ PULV,  song.he@devinci.fr, sophie.depeyre@devinci.fr\\
$^3$ ARCELORMITTAL, AMMR Measurement \& Control, philip.meilland@arcelormittal.com\\
}
\begin{document}
% ---------------------------------------------------------------------
\maketitle
% ---------------------------------------------------------------------

\begin{abstract}
%Des mesures magntiques en ligne ont permis de montrer que lÕon pouvait estimer la rsistance maximale  4\% et la limite lastique  8-10\%. Toutefois, dans le cas de Dual Phase, la seule estimation des proprits mcaniques nÕest pas forcment suffisante pour garantir la qualit du produit. La microstructure conditionne dÕautres caractristiques du matriau difficilement mesurables  lÕaide de simples essais de tractions. Pour que la mesure magntique donne des informations plus prcises sur la microstructure du matriau, il faut pouvoir dcrire suffisamment finement le lien entre la microstructure et les proprits magnto-mcanique en ligne. Dans ce but, ArcelorMittal Maizires Research (AMMR) et et ses partenaire acadmiques conduisent un effort mthodologique de comprhension des liens entre la microstructure et le comportement magntique observable. \\
Le comportement mcanique des aciers dual-phases (DP) est rput fortement sensible  leur microstructure. Un contrle en ligne magntique est envisag. Ce contrle ncessite la mise au point d'une approche inverse incluant  une description fine des liens existant entre la microstructure et les proprits magntiques. Dans le cadre d'une formulation couple magnto-mcanique,  il est ncessaire dÕintroduire une source supplmentaire d'anisotropie  travers le potentiel magnto-mcanique. On prsente une modlisation statique de particule(s) ferromagntique(s) base sur la minimisation d'une fonction d'nergie. Cette modlisation s'appuie sur une mthode du gradient conjugu couple  des mthodes EF permettant la rsolution du problme mcanique.\\

\end{abstract}
\vspace{0.3cm}

\section{Introduction}

\begin{figure}[htbp]
\centering
\subfigure[]{\includegraphics[width=5.5cm]{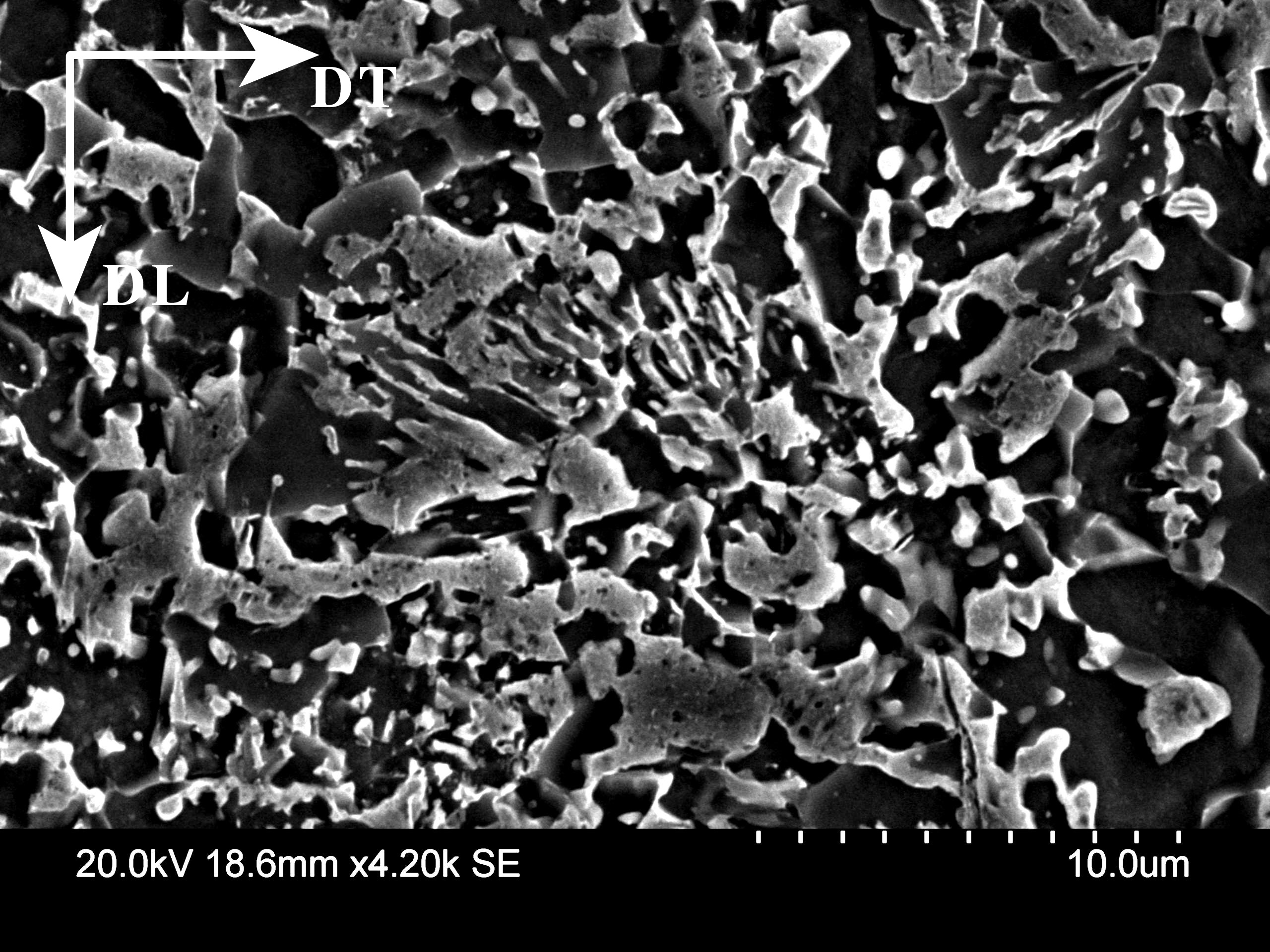}} \quad
\subfigure[]{\includegraphics[width=5.5cm]{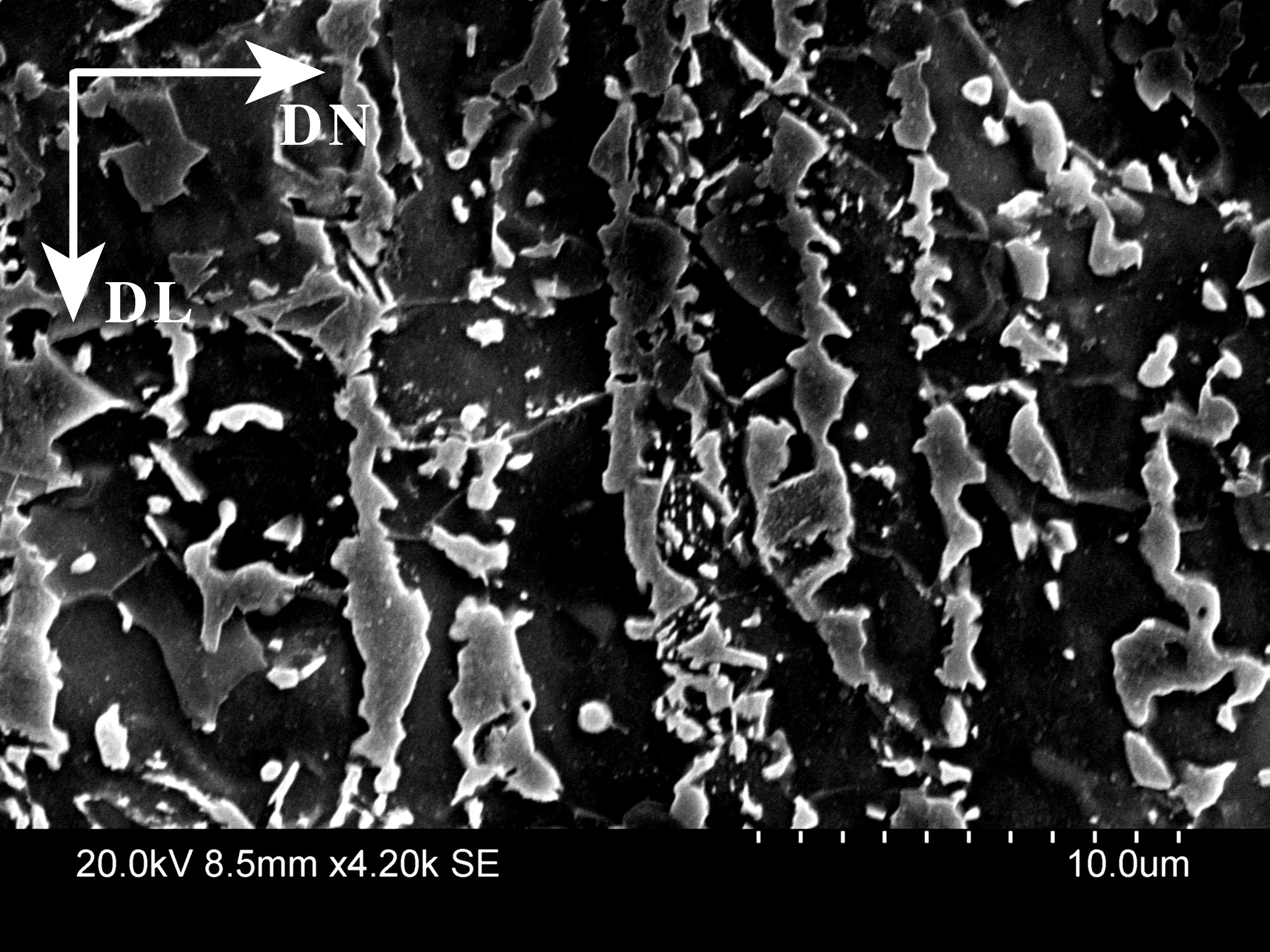}}\quad
\subfigure[reconstruction 3D d'un VER de matire ]{\includegraphics[width=4cm]{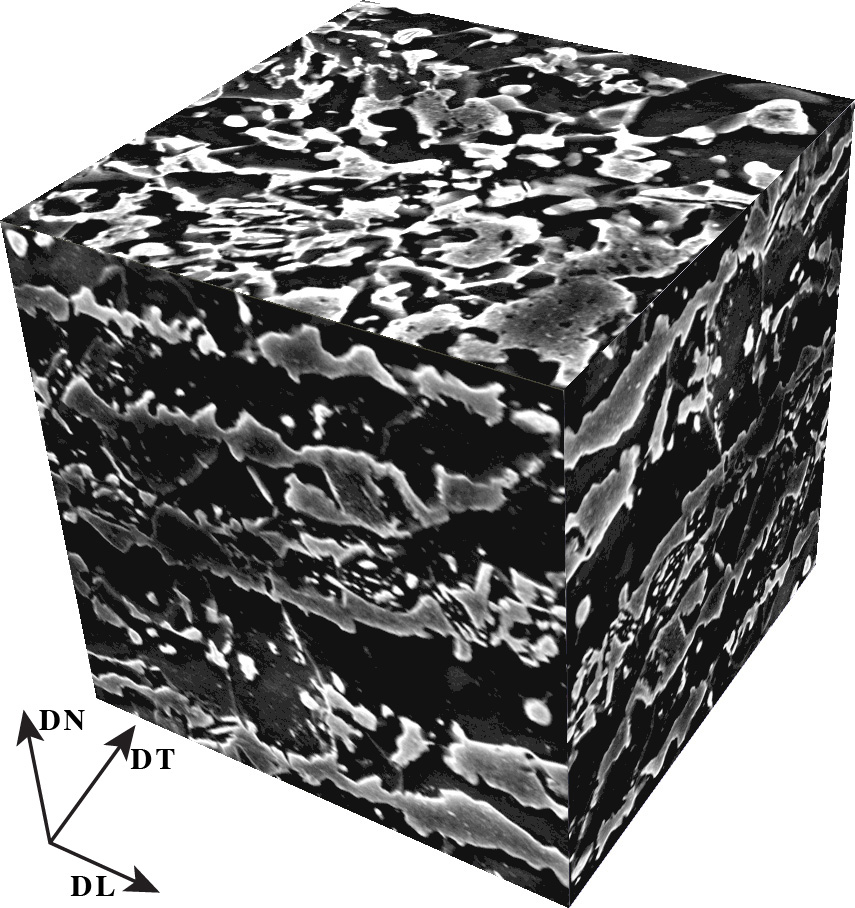}}
\caption{\label{dp} Microstructure d'un acier dual phase (a),(b),(c): martensite en blanc, ferrite en sombre.}
\end{figure}

Les dernires annes ont vues un intrt grandissant des industries automobiles pour l'utilisation d'aciers  haute performance tels que les aciers dual-phases (DP). La production de ces aciers implique plusieurs procds qui conduisent  une microstructure biphase principalement compose d'lots de martensite dure disperss dans une matrice ferritique ductile en proportion variable selon l'histoire thermo-mcanique du matriau (fig \ref{dp}). \\

Les systmes dÕvaluation non destructifs prsentent un intrt particulier pour vrifier lÕhomognit des aciers Dual Phase. Les techniques magntiques ont montr leur efficacit  valuer la microstructure de ces aciers en sortie de recuit/galvanisation. En particulier, le systme 3MA \cite{gabi} est  connu pour son aptitude  valuer de faon non destructive les proprits mcaniques des aciers Dual Phase. Ds 2003, des essais en ligne ont permis de montrer que lÕon pouvait estimer la rsistance maximale avec une erreur infrieure  4\% et la limite lastique  8-10\% prs. Toutefois, dans le cas des Dual Phase, la seule estimation des proprits mcaniques nÕest pas  suffisante pour garantir la qualit du produit. La microstructure conditionne dÕautres caractristiques du matriau difficilement mesurables  lÕaide de simples essais de traction. Ces caractristiques (par exemple lÕtat de contrainte rsiduel) prennent une place de plus en plus importante dans le cahier des charges des clients. Pour que la mesure magntique donne des informations plus prcises sur la microstructure du matriau, il faut pouvoir dcrire suffisamment finement le lien entre la microstructure, l'tat de contrainte et les proprits magntiques en ligne.

\section{Comportement magntique}

LÕapplication dÕun champ magntique aimante les milieux magntiques mais peut galement les dformer. Ce phnomne est appel magntostriction. Il sÕagit dÕune dformation {\itÓspontaneÓ}, intrinsque au matriau. Elle ne dpend que de lÕtat magntique du matriau. Aimantation et magntostriction son lies par une relation non linaire  o\`u $ \boldsymbol{\lambda}$ est le tenseur d'ordre 4 des constantes de magntostriction. 
 
 \begin{equation}
\boldsymbol{\epsilon}^\mu = \boldsymbol{\lambda}:(\vec{m}\otimes \vec{m})
\end{equation}

%\begin{itemize}[label=$\square$]
%
%\item L'aimantation $\vec{m}$ $(A.m^{-1})$ .
%
%\item La magntostriction $\boldsymbol{\epsilon}^\mu$ qui reprsente la dformation subit par un milieu magntique sous l'influence d'un champ magntique extrieur.
%
%\item La susceptibilit $\chi$ ou la permabilit $\mu$  qui caractrisent la nature magntique du matriau (dur, doux, etc\ldots).\\
%
%\end{itemize}

Le mcanisme d'aimantation observable macroscopiquement fut expliqu par Weiss, qui en 1907 mit l'hypothse que le comportement magntique des matriaux magntiques a pour origine l'existence d'une microstructure intragranulaire organise en domaines  portant des directions d'aimantation {\itÊ$\alpha_{_i}$}  diffrentes (\ref{Mdomain}) et spars par des parois (parois de Bloch). Cette interprtation fut confirme exprimentalement lorsque Bitter (1931) observa directement les domaines de Weiss.

%\begin{figure}[htbp]
%\centering
%{\includegraphics[height=3cm]{images_publi/NO.eps}} 
%\subfigure[\label{wall} Dplacement des parois]{\includegraphics[width=5cm]{images_publi/wallmotion.eps}}\qquad
%\subfigure[\label{m_unit} Aimantation locale]{\includegraphics[height=3cm]{images_publi/m.eps}}
%\caption{ \label{weiss} Domaines et parois magntiques: Fe-3\%Si NO (O. Hubert 1998).   }
%\end{figure}

\begin{equation}
\vec{m}_{\alpha}=M_s \begin{pmatrix} \alpha _{_1}&\alpha _{_2}& \alpha _{_3}  \end{pmatrix}
\label{Mdomain}
\end{equation}

 Sous l'influence d'un champ magntique externe, la microstructure magntique volue par dplacement des parois et rotation des moments magntiques. Le dplacement peut tre entrav par la prsence de dfauts structuraux (phases, inclusions, dislocations, joints de grains \ldots). Lorsqu'une paroi rencontre un dfaut, elle se trouve ancre. Le dsancrage  ncessite alors un champ appliqu lgrement suprieur au champ ncessaire au dplacement en absence de dfaut. Les parois progressent alors par sauts brusques et irrversibles. Ces sauts irrversibles sont  l'origine de l'hystrsis du comportement magntique et de sa discontinuit.\\

\`A l'chelle microscopique, la subdivision en domaines rsulte de quilibre nergtique entre diffrentes contributions lmentaires. Trois contributions  en particulier sont  l'origine de cette subdivision, l'nergie d'change, l'nergie d'anisotropie et l'interaction dipolaire \cite{schaffer}. L'interaction d'change traduit la tendance des moments magntiques voisins  s'aligner dans la mme direction. Ce phnomne se traduit  l'chelle macroscopique par un un gradient d'aimantation correspondant  une force de rappel lastique. L'interaction d'change est une interaction  purement locale qui dcroit rapidement avec la distance, et ne dpend que des proprits microscopiques du matriau. On y associe usuellement un potentiel de la forme:

\begin{equation}
\mathcal{E}_{ech}=\mu_0\  \boldsymbol{A}. \| \vec{\nabla} \vec{m}\| ^2
\end{equation}

L'anisotropie reprsente l'interaction des moments magntiques avec le potentiel cr par la symtrie du cristal. Le rsultat d'une telle interaction est la prfrence des moments magntiques  s'orienter dans certaines directions nergtiquement favorables  (direction de symtrie du cristal). Cette interaction prend la forme d'une fonction  $\Psi(\vec{d},\vec{m})$ paire s'annulant suivant les diffrents axes d'anisotropie ($\vec{d}$) du cristal considr.

\begin{equation}
\mathcal{E}_{a}= \Psi(\vec{d}.\vec{m})
\end{equation}

Ainsi dans le cas d'une symtrie cubique,  la fonction $\Psi(\vec{d},\vec{m})$ prend la forme

\begin{equation}
\mathcal{E}_{a}=K_0 + K_1(\alpha_1^2\alpha_2^2 + \alpha_2^2\alpha_3^2 + \alpha_1^2\alpha_3^2)+K_2(\alpha_1^2\alpha_2^2\alpha_3^2)
\end{equation}

%\begin{figure}[ht!]
% \centering
%\includegraphics[width=5cm]{images_publi/cubicAnis}
%\caption{\label{cubic} Energie d'anisotropie cubique.}
%\end{figure}

L'interaction dipolaire joue un rle fondamental dans la formation des domaines. Celle-ci est lie  l'interaction mutuelle des moments magntiques et  la gomtrie du milieu. L'aimantation cre dans un milieu par l'action d'un champ magntique extrieur gnre en retour un champ. Ce champ  appel champ dmagntisant $\vec{H}_d({\vec{m}})$  satisfait les quations de Maxwell $(\vec{\nabla} \wedge \vec{H}_d({\vec{m}}) = \vec{0})$ et donc drive d'un potentiel scalaire $\zeta$ vrifiant l'ensemble des quations (\ref{Hd}) et (\ref{poisson}).

\begin{equation}
\vec{H}_d({\vec{m}})=-\vec{\nabla} \zeta \label{Hd}
\end{equation}

\begin{equation}
\Delta \zeta(\vec{r}) = \frac{1}{\mu_0}\boldsymbol{\nabla}\vec{m}(\vec{r}) \  \forall\ \vec{r}\in \Omega \quad ;\quad
\Delta \zeta(\vec{r}) =0\   \forall\ \vec{r}\in\mathbb{R}-\Omega \quad ;\quad
\left [\zeta \right] =0\  \forall\ \vec{r}\in\partial\Omega\\
\label{poisson}
\end{equation}

Le potentiel $\zeta$ est solution du problme de Poisson  (\ref{poisson}) et traduit le caractre non local du champ dmagntisant. Ce champ conditionne un grand nombre de microstructures magntiques observes exprimentalement. Du fait de son caractre non local, le calcul de cette grandeur est en gnral trs coteux en temps de calcul. On lui associe usuellement un potentiel de la forme:

\begin{equation}
\mathcal{E}_{d}=-\frac{\mu_0}{2}  \vec{H}_d(\vec{m}).\vec{m}
\end{equation}

Enfin la dernire interaction, traduit l'influence du champ appliqu sur les moments magntiques. Il s'agit de l'nergie de Zeemann.
\begin{equation}
\mathcal{E}_{h}=-\mu_0\ \vec{H}_{ext}.\vec{m}
\end{equation}

\section{Modlisation Micromagntique d'une particule ferromagntique}

Le micromagntisme est une approche thorique permettant de dcrire le processus d'aimantation  une chelle suffisamment large pour remplacer les moments magntiques atomiques par des fonctions continues, et suffisamment fine pour rendre compte des zones de transition entre domaines magntiques \cite{brown}\cite{aharoni}. Cette approche continue de la thorie du ferromagntisme fait le lien entre la description quantique de la structure des spins et la thorie de l'lectromagntisme de Maxwell o\`u les proprits magntiques du matriau sont dcrites par des constantes volumiques homognes (susceptibilit $\chi$ et permabilit $\mu$). 

\subsection{Micromagntisme}

Il s'agit  d'une approximation  base sur la minimisation d'une fonctionnelle nergtique . Cette minimisation par rapport  l'aimantation donne lieu  un tat d'quilibre mtastable associ  une configuration particulire de la structure magntique. La contribution apporte par Brown \cite{brown} fut de dfinir une expression de l'nergie libre $\mathcal{E}_{tot}$ comme somme des contributions internes et externes nonces prcdemment. 

\begin{equation}
\mathcal{ E}_{tot}(\vec{m})=\int_{_\Omega} \mathcal{E}_{h}+\mathcal{E}_{ex}+\mathcal{E}_{a}+\mathcal{E}_{d}\ d\Omega
\end{equation} 

A l'quilibre, on doit alors satisfaire la relation de stationnarit nergtique:

 \begin{equation}
 \vec{m}(\vec{x}) =   Min_{({\vec{m}\in\mathbb{R}^n})}\left(  \mathcal{ E}_{tot}(\vec{m} )\right) \qquad \forall \vec{x}\ \in\ \Omega
 \label{pbMin}
\end{equation}

 Pour tout champ appliqu $\vec{H}_{ext}$,  $M_s$ donn, on cherche une solution $\vec{m}$ sous la contrainte
\begin{equation}
 \| \vec{m}\| =M_s \ \forall \vec{x}\ \in\ \Omega
 \label{constrain}
\end{equation}
Pour une petite variation de l'aimantation locale $\delta\vec{m}$, la variation de l'nergie libre totale  est donne par: 
\begin{equation}
\delta\mathcal{E}_{tot}(\vec{m})=\mathcal{E}_{tot}(\vec{m}+\delta \vec{m})-
\mathcal{E}_{tot}(\vec{m})=0
\label{minmicro}
\end{equation}
Soit:
\begin{equation}
\delta\mathcal{E}_{tot}(\vec{m}) \approx \int_{_\Omega} \left(  -\mu_o\vec{H}_{ext} - \mu_o \vec{H}_d+\frac{\partial \Psi(\vec{d}.\vec{m})}{\partial \vec{m} } +  \mu_o \boldsymbol{A} \Delta\vec{m} \right).\delta \vec{m} \ d\Omega + \oint  \mu_o \boldsymbol{A} \frac{\partial \vec{m} }{\partial \vec{n}}.\delta \vec{m}\ \  \partial\Omega
\end{equation}

O $\vec{n}$ est le vecteur unitaire normal  la surface qui dlimite le volume $\Omega$. La condition de stationnarit de l'nergie totale peut donc se mettre sous la forme suivante:

\begin{equation}
\delta\mathcal{E}_{tot}(\vec{m}) = - \int_{_\Omega} \mu_o\vec{H}_{eff}.\delta \vec{m} \ d\Omega + \oint  \mu_o \textbf{A} \frac{\partial \vec{m} }{\partial \vec{n}}.\delta \vec{m} \ \  \partial\Omega =0 \label{vari}
\end{equation}

\begin{equation}
\vec{H}_{eff}=\vec{H}_{ext}+\vec{H}_d - \boldsymbol{A} \Delta\vec{m} - \frac{1}{2}\frac{\partial \Psi(\vec{d}.\vec{m})}{\partial \vec{m}}  = -\frac{1}{\mu_o}\frac{\partial \mathcal{E}_{tot}(\vec{m})}{\partial \vec{m}} \label{Heff}
\end{equation}

Le champ effectif $\vec{H}_{eff}$ dfini par (\ref{Heff}) est le champ  localement ressenti par chaque moment magntique. De l'galit (\ref{vari}) on tire un systme appel  $"$systme de Brown$"$ \cite{brown} qui dfini deux conditions  respecter  l'quilibre. Une configuration magntique qui  ralise le minimum nergtique doit  vrifier le systme d'quation de Brown :

\begin{equation}
\begin{matrix}
\vec{m}\wedge\vec{H}_{eff}&=\vec{0}&\forall\quad \vec{x}\ \in\ \Omega \qquad
\vec{m}\wedge\frac{\partial\vec{m}}{\partial\vec{n}}&=\vec{0}&\forall\quad \vec{x}\ \in\ \partial\Omega
\end{matrix}
 \label{bbrow}\end{equation}

\subsection{Choix de rsolution}

Notre approche du problme d'optimisation (\ref{pbMin}) combine une mthode itrative (gradient conjugu) et une  mthode lments finis permettant la description du milieu. Le gradient conjugu (algorithme \ref{algo1} dtaill ci-dessous) permet en gnral de traiter des problmes d'optimisation. On peut montrer que cette mthode applique  une fonctionnelle quadratique elliptique converge en $n$ itrations. Cependant, dans le cas du micromagntisme, la contrainte (\ref{constrain}) impose  l'aimantation locale associe   la non convexit de l'nergie magntocristalline n'assurent pas l'unicit de la solution.

\begin{multicols}{2}
\begin{algorithm}[H] 
Suppos: $\vec{m}_0$,  $ \delta\mathcal{W}_{tot}(\vec{m}_0)= \vec{g}_0$,  $\vec{w}_0=\vec{g}_0$ \;
\While{$n\geq 1$, }{
Trouver $\rho_n\in \mathbb{R}$ Tel que $\forall$ $\rho\in \mathbb{R}$\; 
$\mathcal{W}_{tot}(\vec{m}_n-\rho_{_n}\vec{w}_n)<\mathcal{W}_{tot}(\vec{m}_n-\rho \vec{w}_n)$ \;
$\vec{m}_{n+1}=\vec{m}_n-\rho_{_n}\vec{w}_n$\;
$\vec{g}_{n+1}= \vec{H}_{eff}(\vec{m}_{n+1}) $\;
$\vec{w}_{n+1}=\vec{g}_{n+1}+r_n\vec{w}_n$\;
O\`u l'on choisit $r_n = \frac{\vec{g}_{n+1}.(\vec{g}_{n+1}-\vec{g}_n)}{\| \vec{g}_n \|^2_{_{\mathbb{R}^n}}}$\;}
\caption{\label{algo1}Gradient conjugu.}
\end{algorithm}

\begin{algorithm}[H] 
Suppos: $\vec{m}_0$,  $ \delta\mathcal{W}_{tot}(\vec{m}_0)= \vec{g}_0$,  $\vec{w}_0=\vec{g}_0$ \;
\While{$n\geq 1$, }{
Trouver $\rho_n\in \mathbb{R}$ Tel que $\forall$ $\rho\in \mathbb{R}$\; 
\textcolor{red}{$\mathcal{W}_{tot}\left( \frac{\vec{m}_n-\rho_{_n}\vec{w}_n}{\| \vec{m}_n-\rho_{_n}\vec{w}_n\|} \right)<\mathcal{W}_{tot}\left( \frac{\vec{m}_n-\rho \vec{w}_n}{\| \vec{m}_n-\rho \vec{w}_n\|} \right)$}\;

\textcolor{red}{$\vec{m}_{n+1}= \frac{\vec{m}_n-\rho_{_n}\vec{w}_n}{\|\vec{m}_n-\rho_{_n}\vec{w}_n\|}$}\;
$\vec{g}_{n+1}= \vec{H}_{eff}(\vec{m}_{n+1}) $  \;
$\vec{w}_{n+1}=\vec{g}_{n+1}+r_n\vec{w}_n$\;
O\`u l'on choisit $r_n = \frac{\vec{g}_{n+1}.(\vec{g}_{n+1}-\vec{g}_n)}{\| \vec{g}_n \|^2_{_{\mathbb{R}^n}}}$\;
} 
\caption{\label{algomicro}Gradient conjugu avec la contrainte: $\| \vec{m}\|=M_s$}
\end{algorithm}
\end{multicols}

 Le problme mathmatique d'existence et d'unicit de la solution reste encore  ce jour un problme ouvert \cite{song2}. Mais, en supposant l'existence d'une solution, la problmatique est alors de construire un schma numrique conservant au mieux les proprits du systme  savoir: Les proprits du champ dmagntisant (\ref{poisson}), la dcroissance de l'nergie, et la conservation de la norme locale de l'aimantion. Cette dernire remarque nous conduit  effectuer un changement de variable dans la dfinition de l'nergie, et dans la construction de la suite $\vec{m}_{n+1}$. 

\subsection{Couplage magnto-mcanique}

La variation d'aimantation d'un milieu magntique (magntostrictif) induit une variation de volume  $\boldsymbol{\epsilon}^\mu$ appele magntostriction spontane.  Celle-ci correspond en gnral  une incompatibilit ({\it i.e. dformation ne drivant pas d'un champ de dplacement }). La dformation lastique $\boldsymbol{\epsilon}^e$ du milieu aimant doit alors corriger  l'incompatibilit de la magntostriction, ce qui se traduit par un tat de contrainte non nul, mme en l'absence de toute sollicitation mcanique externe. On s'intresse ici  l'volution d'un systme mcanique qui, sous l'action d'une sollicitation interne et ou externe, volue d'un tat d'quilibre  un autre. Lorsque la magntostriction spontane $\boldsymbol{\epsilon}^\mu$ du milieu est connue, la dformation magnto-lastique correspondante $\boldsymbol{\epsilon}$ est obtenue par additivit $(\boldsymbol{\epsilon} = \boldsymbol{\epsilon}^\mu + \boldsymbol{\epsilon}^e)$, les dformations considres rentrant dans le cadre de l'hypothse des petites perturbations. La dformation magnto-lastique totate $\boldsymbol{\epsilon}$, drivant d'un champ de dplacement $\vec{u}$  (\ref{gradeps}), et la contrainte $\boldsymbol{\sigma}$ obissant  l'quation d'quilibre, les champs mcaniques vrifient les quations suivantes:
  
\begin{equation}
\begin{array}{clr}
\boldsymbol{\boldsymbol{\epsilon}}=&\frac{1}{2} \left( \boldsymbol{\nabla}\vec{u} + \boldsymbol{\nabla}^t\vec{u}\right)&\ sur\ \Omega \\ 
\end{array}\label{gradeps}
\end{equation}

\begin{equation}
\vec{\nabla}.\boldsymbol{\sigma}=\ 0\ sur \ \Omega \qquad
\boldsymbol{\sigma} =\boldsymbol{C} :\boldsymbol{\epsilon}^e \ sur \ \Omega 
\label{meca}
\end{equation}

o $\boldsymbol{C}$ dsigne le tenseur  d'lasticit d'ordre 4 du milieu. Les conditions aux limites du problme sont donnes par:

\begin{equation}
\boldsymbol{\sigma}.\vec{n}=\vec{T}_d\ sur\ \partial\Omega_t\qquad
\vec{u}=\vec{u}_d\ sur\ \partial\Omega_u
\label{boundcond}
\end{equation}

La linarit des dformations permet de reformuler l'quilibre mcanique  de manire  introduire explicitement  une sollicitation interne d'origine magntique (magntostrictive).

\begin{equation}
\vec{\nabla}.{\boldsymbol{\sigma}}^* - \vec{f}^\mu = 0 \label{forte}
\end{equation}

O\`u $\boldsymbol{\sigma}^*  $ reprsente la contrainte totale et $ \vec{f}^\mu = \vec{\nabla}.\left( \boldsymbol{C}:\boldsymbol{\epsilon}^\mu\right) $  la densit de forces d'origine magntostrictive, consquence directe de l'tat d'aimantation du matriau. \`A l'image du micromagntisme le problme mcanique peut galement se rduire  un problme d'optimisation o\`u le champ de dplacement minimise un potentiel donn par l'application d'une formulation variationnelle au problme (\ref{forte}):

\begin{equation}
 \vec{u} =     Min_{(\vec{v}\in\mathcal{H}^1_\Omega)}\  \mathcal{ E}_{\boldsymbol{\sigma}}(\vec{v})\qquad avec\    \vec{v}=u_d \quad sur \quad \  \partial\Omega_d \label{Minmeca}
 \end{equation}
 
 Avec: 
 
\begin{equation}
\mathcal{ E}_{\boldsymbol{\sigma}^*}(\vec{v})=\frac{1}{2} \boldsymbol{\epsilon}(\vec{v}):\boldsymbol{C}:\boldsymbol{\epsilon} (\vec{v})\ -\ \boldsymbol{\epsilon}(\vec{v}):\boldsymbol{C}:\boldsymbol{\epsilon}^\mu\  
\end{equation} 
 
 Dans le cas d'une modlisation type FEM base sur approche en dplacement, le problme de minimisation (\ref{Minmeca}) peut se simplifier sous la forme du systme linaire (\ref{linearsyst}) o\`u $\mathbb{K}$ ,  $U$ et $F$ reprsentent respectivement la matrice de rigidit du systme, le vecteur dplacement gnralis et le vecteur force gnralis traduisant les efforts d'origine magntostrictive. 
\begin{equation}
\mathbb{K}.{U}= F \label{linearsyst}
\end{equation} 
 L'expression approche  l'aide d'lments finis isoparamtriques  sur le domaine discrtis $\Omega_h$ est donne pour des fonctions de forme $\phi$ linaires:
\begin{equation} 
\mathbb{K}_h=\sum_{i=1}^{N_h}\sum_{j=1}^{N_h}\int_{\Omega_h} {\vec{\nabla}\phi_j:\boldsymbol{C}:\vec{\nabla}\phi_i} \ d\Omega_h \qquad F_h =\sum_{j=1}^{N_h} \int_{\Omega_h}{\boldsymbol{\epsilon}^\mu_h:\boldsymbol{C}:\vec{\nabla}\phi_j}\ d\Omega_h \qquad {U}_h = \sum_{j=1}^{N_h} u_j
 \end{equation}

Dans le cadre d'une approche couple, les contributions lies  l'quilibre d'un milieu magntique, mais aussi les contributions lies  l'quilibre d'un milieu dformable et les diffrentes  interactions entre ces deux phnomnes participent  l'quilibre nergtique. L' nergie libre dcrivant au mieux un milieu magntique dformable prend alors la forme suivante:

\begin{equation}
\mathcal{ E}_{tot}(\vec{m},\vec{u})=\mathcal{E}_{z}+\mathcal{E}_{ex}+\mathcal{E}_{a}+\mathcal{E}_{d}+\textcolor{red}{\mathcal{E}_{\boldsymbol{\sigma}}}
\end{equation} 

La condition de stabilit de l'nergie est alors obtenue si et seulement si les champs d'aimantation et de dplacement minimisent simultanment l'nergie libre totale. Cette condition s'crit :

\begin{equation} 
\delta\mathcal{ E}_{tot}=\frac{\partial \mathcal{ E}_{tot}}{\partial \vec{m}}\delta \vec{m}+\frac{\partial \mathcal{E}_{tot}}{\partial \vec{u}}\delta \vec{u}=0 \label{coupled}
\end{equation}

Sous les contraintes  (\ref{constrain}) , (\ref{boundcond}). Une condition de minimisation est l'annulation de toutes les drives partielles, ce qui permet la rsolution du problme magntique et du problme mcanique sparment. On a alors  rsoudre le systme.

\begin{equation}
\frac{\partial \mathcal{ E}_{tot}}{\partial \vec{m}}\  =\vec{0}\qquad
\frac{\partial \mathcal{ E}_{tot}}{\partial \vec{u}}\ =\vec{0}\qquad
\forall\ \vec{x}\ \in\ \Omega \label{dev_u}
\end{equation}

La premire minimisation fournit toujours la condition de couple donne par l'quation (\ref{bbrow}). La seconde minimisation fournit toujours le systme linaire donn par l'quation (\ref{linearsyst}). La subtilit cependant rside dans la dfinition du champ effectif, qui se voit complte par un terme supplmentaire appel champ d'anisotropie lastique induit, et li  la drivation de l'nergie magnto-lastique par rapport  l'aimantion tel que: 

\begin{equation}
\vec{H}_{eff}=\vec{H}_{ext}+\vec{H}_d - \boldsymbol{A} \Delta\vec{m} - \frac{1}{2}\frac{\partial \Psi(\vec{d}.\vec{m})}{\partial \vec{m}} -  \textcolor{red}{\vec{H}_{\sigma}} = -\frac{1}{\mu_o}\frac{\partial \mathcal{E}_{tot}(\vec{m})}{\partial \vec{m}}
\end{equation} 

Avec 
\begin{equation}
\vec{H}_{\sigma} =-\frac{1}{\mu_o}\frac{\partial\mathcal{E}_{\boldsymbol{\sigma}}}{\partial \vec{m}}   =-\frac{1}{\mu_o}\frac{\partial \left( - \boldsymbol{\sigma}:\boldsymbol{\epsilon}^\mu \right)}{\partial \vec{m}}
\end{equation}

Dans le cas d'un matriau  symtrie cubique et en ngligeant la magntostriction de volume (hypothse usuelle pour les matriaux magntiques mtalliques), le tenseur de magntostriction prend la forme simple:

\begin{equation}
\boldsymbol{\epsilon}^\mu_{ii} = \frac{3}{2}\lambda_{100}(\alpha_i^2 - \frac{1}{3}) \quad , \quad  \boldsymbol{\epsilon}^\mu_{i \neq j} = \frac{3}{2}\lambda_{111}\alpha_i\alpha_j
\end{equation} 
L'nergie $"$magnto-lastique$"$ $(\boldsymbol{\sigma}:\boldsymbol{\epsilon}^\mu)$ se dveloppe sous la forme:

\begin{equation}
\boldsymbol{\sigma}:\boldsymbol{\epsilon}^\mu = \sum_i\boldsymbol{\sigma}_{ii}\boldsymbol{\epsilon}^\mu_{ii} + 2\sum_{j\neq i}\boldsymbol{\sigma}_{ij}\boldsymbol{\epsilon}^\mu_{ij} = 
\frac{3}{2}\lambda_{100} \sum_i\boldsymbol{\sigma}_{ii} (\alpha_i^2 - \frac{1}{3}) + 3\lambda_{111}\sum_{j\neq i}\boldsymbol{\sigma}_{ij}\alpha_i\alpha_j
\end{equation} 

et on exprime le champ d'anisotropie lastique :

\begin{equation}
\vec{H}_{\boldsymbol{\sigma} _ i} = \frac{3}{\mu_0}(\lambda_{100} \boldsymbol{\sigma}_{ii} \alpha_i + \lambda_{111}\sum_{j\neq i}\boldsymbol{\sigma}_{ij}\alpha_j)
\end{equation} 

La contribution supplmentaire au champ effectif ncessite  chaque itration de connatre localement l'tat de contrainte mcanique par la rsolution des quations d'quilibre (\ref{meca}) et (\ref{boundcond}). Cette rsolution se rsume  l'inversion du systme linaire (\ref{linearsyst}). Nous remplaons alors l'agorithme \ref{algomicro} par l'algorithme \ref{algomicromeca} o\`u la rsolution du problme mcanique  chaque itration permet de redfinir l'tat de contrainte. Il s'agit de l'approche dite {\it"constrained"}, qui diffre de l'approche {\it"relaxed"} o\`u l'on nglige l'effet de la contrainte intrinsque, en remplaant la dformation totale par la magntostriction $(\boldsymbol{\epsilon}\approx\boldsymbol{\epsilon}^\mu)$. Cette approximation trs rpandue dans le cadre du couplage magnto-mcanique soulve le problme de la compatibilit des dformations et de  la  configuration des spins obtenue \cite{yi}.\\

\begin{algorithm}[H] 
Suppos: $\vec{m}_0$ , $ \delta\mathcal{E}_{tot}(\vec{m}_0)= \vec{g}_0$,  $\vec{w}_0=\vec{g}_0$, \textcolor{red}{$\mathbb{K}_h.U_{h_0}=F_{h_0}$}\;
\While{$n\geq 1$, }{
Trouver $\rho_n\in \mathbb{R}$ Tel que $\forall$ $\rho\in \mathbb{R}$\; 
$\mathcal{E}_{tot}\left( \frac{\vec{m}_n-\rho_{_n}\vec{w}_n}{\| \vec{m}_n-\rho_{_n}\vec{w}_n\|},\textcolor{red}{\boldsymbol{\sigma}_n} \right)<\mathcal{E}_{tot}\left( \frac{\vec{m}_n-\rho \vec{w}_n}{\| \vec{m}_n-\rho \vec{w}_n\|},\textcolor{red}{\boldsymbol{\sigma}_n} \right)$\;

$\vec{m}_{n+1}= \frac{\vec{m}_n-\rho_{_n}\vec{w}_n}{\|\vec{m}_n-\rho_{_n}\vec{w}_n\|}$\;
\textcolor{red}{$\mathbb{K}_h.U_{h_{n+1}}=F_{h_{n+1}}$}\;

$\vec{g}_{n+1}=$ \textcolor{red}{$\vec{H}_{eff}(\vec{m}_{n+1},\boldsymbol{\sigma}_{n+1})$}\;
$\vec{w}_{n+1}=\vec{g}_{n+1}+r_n\vec{w}_n$\;
O\`u l'on choisit $r_n = \frac{\vec{g}_{n+1}.(\vec{g}_{n+1}-\vec{g}_n)}{\| \vec{g}_n \|^2_{_{\mathbb{R}^n}}}$\;
} 
\caption{\label{algomicromeca}Gradient conjugu Magnto-Mcanique: approche {\it"contrainte"}.}

\end{algorithm}
\vspace{0.3cm}

\section{Applications}

Le code de calcul micromagntique statique 2D {\it"sivimm2d"} \cite{song}  nous a  servi de base  l'implmentation de la formulation magnto-mcanique (algorithme \ref{algomicromeca}). Nous tudions l'volution d'un monocristal ferromagntique  symtrie cubique dont les constantes magntique et mcanique sont rsumes dans le tableau (\ref{constantes}).  Du fait de la restriction 2D, l'hypothse de contrainte plane $(\sigma_{zz}=0)$ a t retenue, ce qui ncessite la redfinition du tenseur d'lasticit $\boldsymbol{C}$  initialement 3D par un tenseur plan $\boldsymbol{C}_{Plan}$ tenant compte de cette restriction.
\begin{figure}[ht!]
 \centering
\subfigure[Bords libres]{\includegraphics[width=3.5cm]{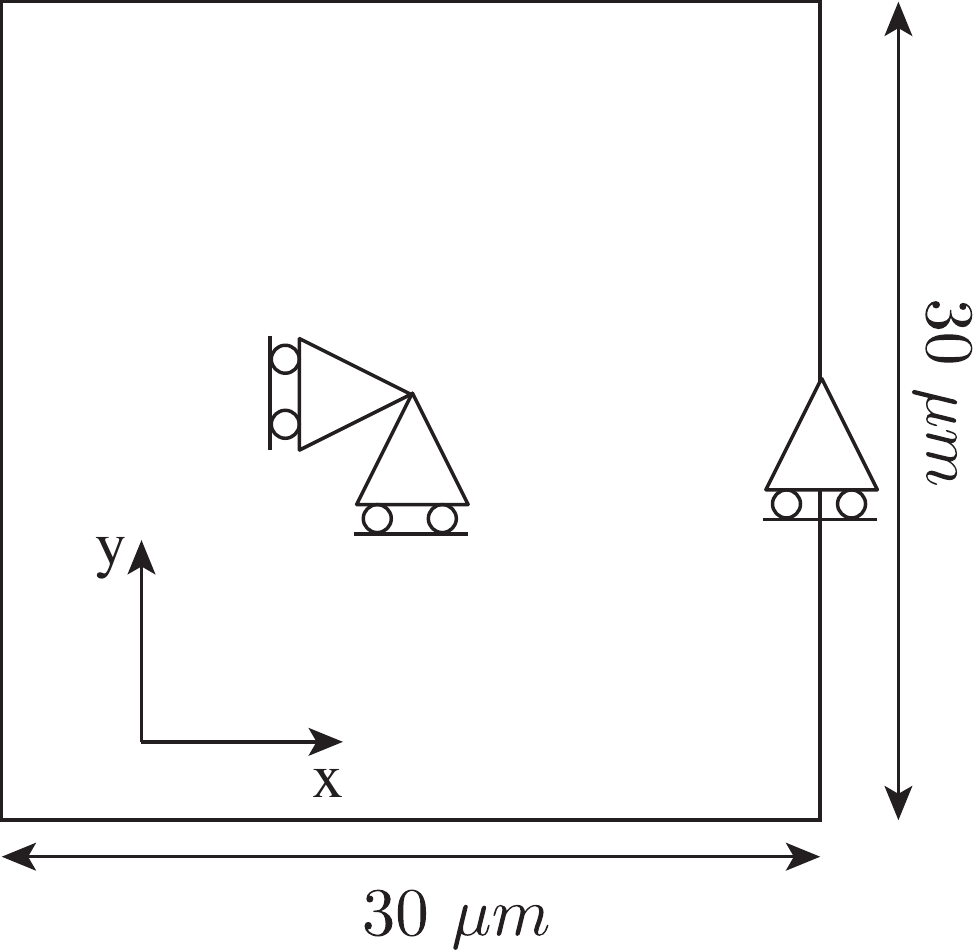}}\qquad
\subfigure[\label{cl_biax} Traction ]{\includegraphics[width=3cm]{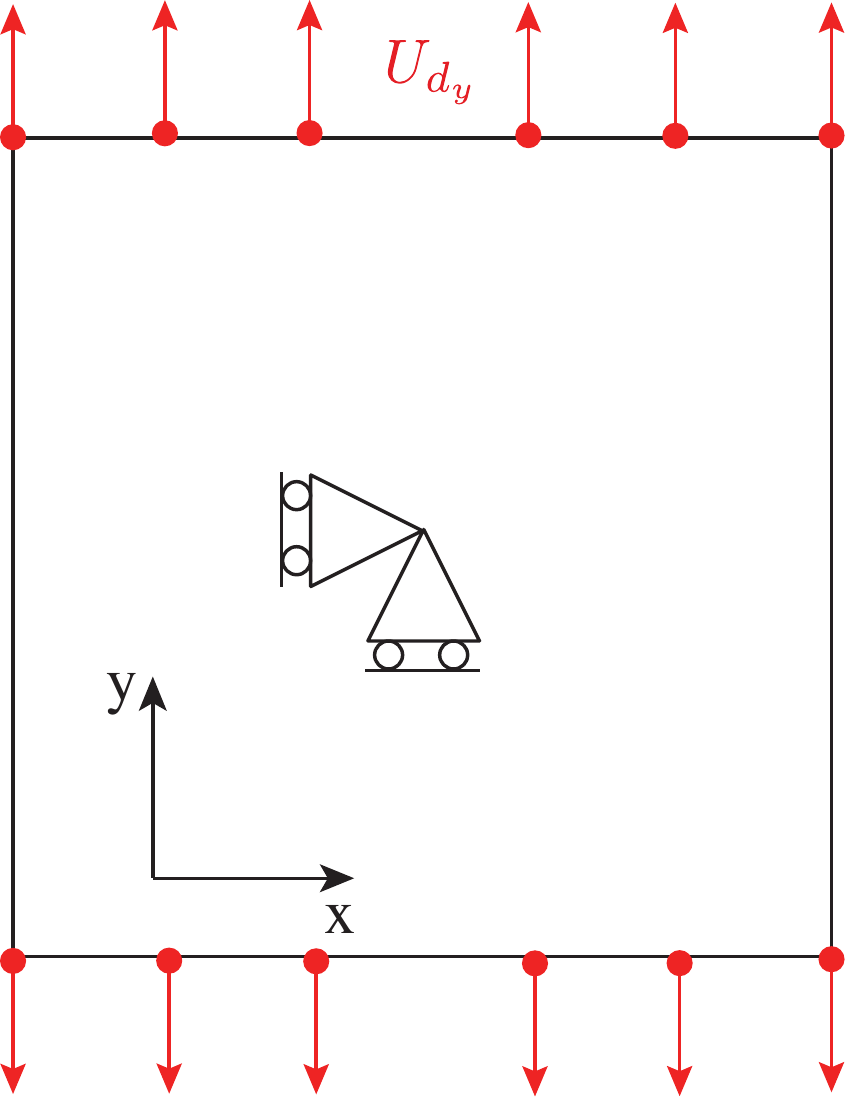}}
\caption{\label{milieu} Domaine d'tude et condition aux limites mcaniques ($U_{d_y}=\pm 0,5\mu m$).}
\end{figure}
\begin{table}[ht!]
   \centering
   \begin{tabular}{llll}
$Ms  = 1,7\ .10^6\ A.m^{-1}$ & $a   = 1,0105\ .10^4\ J.m^{-1} $ & $K_1 = 48\ .10^3\ J.m^{-3}$ &  $K_2 = 0$ \\
 $\lambda_{100} = 21\ .10^{-6}$    &$\lambda_{111} =-21\ .10^{-6}$ &2 axes d'anisotropie $[100]=\vec{x}$ et $[010]=\vec{y}$\\
$C_{11} = 228\ GPa$ & $C_{12} = 132\ GPa$ & $C_{44} = 116.5\ GPa$
\end{tabular}
\caption{\label{constantes} Constantes physiques (monocristallines) utilises pour les simulations}
\end{table}

\begin{figure}[ht!]
 \centering
\subfigure[\label{compa} influence du couplage magnto-mcanique $(\vec{m}.\vec{x})$]{\includegraphics[width=9cm]{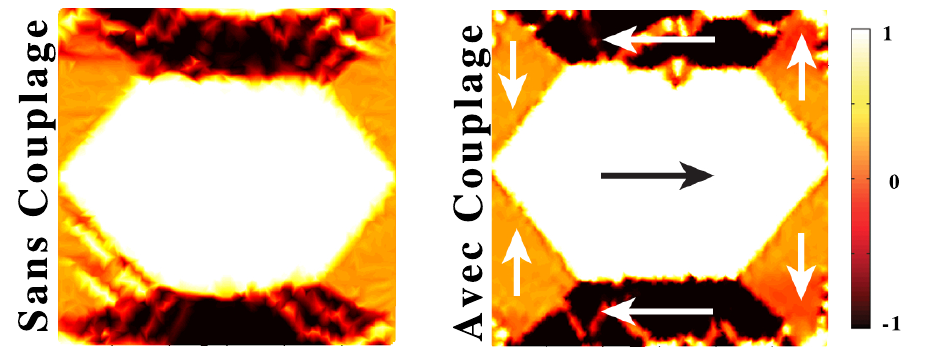}}\quad
\subfigure[\label{SIG} $\frac{1}{2}\boldsymbol{\sigma}:\boldsymbol{\epsilon}^e\  (J.m^{-3})$]{\includegraphics[width=4cm]{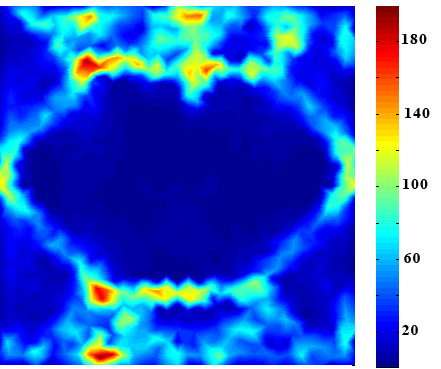}}
\caption{ Influence du couplage (a) et nergie lastique (b) associe  une configuration $(\vec{H}_{ext}=0.15M_s\ \Vec{x})$}
\end{figure}
Les premires simulations sur le monocristal ont permis de mettre en vidence l'apport du couplage magnto-mcanique (fig. \ref{compa}): En effet, en absence d'effet magnto-lastique, la microstructure magntique est domine par les effets dmagntisant d'anisotropie qui tendent  aligner les moments magntiques suivant les axes faciles $<100>$ et  favoriser l'apparition de domaines de refermeture du flux prs des frontires de manire  minimiser l'nergie magntostatique. La contrainte $\boldsymbol{\sigma}$ et son nergie associe agissent comme une source supplmentaire d'anisotropie $(K_{\boldsymbol{\sigma}})$, et en absence de sollicitation mcanique extrieure, celle ci agit sur la taille caractristique des parois magntiques dfinies respectivement par:
\begin{equation}
l_{K} \propto \ \sqrt{\frac{a}{K}} \quad , \quad l_{\boldsymbol{\sigma}} \propto \ \sqrt{\frac{a}{K+K_{\boldsymbol{\sigma}}}}
\end{equation}

L'apport nergtique d  l'tat de contrainte intrinsque (fig. \ref{SIG}) a donc pour effet de stabiliser la microstructure magntique tel qu'on peut le constater (fig. \ref{compa}) \\ 
%\begin{figure}[ht!]
% \centering
%% \subfigure[$\sigma_{xx}$ (MPa)]{\includegraphics[width=5cm]{images_publi/S11}} \ 
%%\subfigure[$\sigma_{yy}$ (MPa)]{\includegraphics[width=5cm]{images_publi/S22}} \ 
%%\subfigure[$\sigma_{xy}$ (MPa)]{\includegraphics[width=5cm]{images_publi/S12}}\ 
%% \subfigure[\label{edef} $\frac{1}{2}\boldsymbol{\sigma}:\boldsymbol{\epsilon}^e\quad (J.m^{-3})$]{\includegraphics[width=5cm]{images_publi/EnerE}} 
%\includegraphics[width=4cm]{images_publi/EnerE}
%\caption{\label{SIG} \'Energie lastique $(\frac{1}{2}\boldsymbol{\sigma}:\boldsymbol{\epsilon}^e\  (J.m^{-3}))$ associe  une configuration magntique  champ $\vec{H}_{ext}=\vec{0}$.  }
%\end{figure}

Le couplage magnto mcanique introduit prcdemment laisse la possibilit d'explorer l'influence de l'tat de chargement mcanique externe sur la configuration des moments magntiques. Ainsi sur le mme domaine d'tude, on choisit,  champ fix, d'appliquer une sollicitation mcanique perpendiculaire au champ appliqu (fig. \ref{cl_biax}) et d'observer l'volution de la microstructure magntique sous l'action de ce chargement. $\Sigma_{yy}$ dfinit la contrainte moyenne applique correspondant au dplacement impos.

% Sur le mme domaine d'tude, un ensemble de sollicitations mcaniques est appliqu (fig. \ref{cl_biax}) de manire  dfinir le plan de chargement mcanique $(\boldsymbol{\sigma}_{xx},\boldsymbol{\sigma}_{yy})$ (fig. \ref{stress_plan}). Ces sollicitations sont appliques sous forme de dplacement impos sur les frontires du domaine, et la contrainte exprime est un moyenne sur l'ensemble du milieu. On superpose  ce chargement mcanique une sollicitation magntique autour d'une valeur  l'quilibre de $H_{ext}=0.15\ M_s$ parallle  l'axe d'anisotropie du matriau.

\begin{figure}[ht!]
 \centering
%   \subfigure[\label{init}$\vec{H}_{ext}=0,15\ M_s\ .\vec{x}$]{\includegraphics[width=4cm]{images_publi/MxSignull.pdf}}\\
%  \subfigure[\label{cl_biax} $U_{d_y}=\pm 0,5$]{\includegraphics[width=4cm]{images_publi/Cl_biax.pdf}}\\
  \subfigure[\label{pSyy} $\Sigma_{yy}=4,37\ Gpa\quad \Sigma_{xx}=\Sigma_{xy}=0$]{\includegraphics[width=7cm]{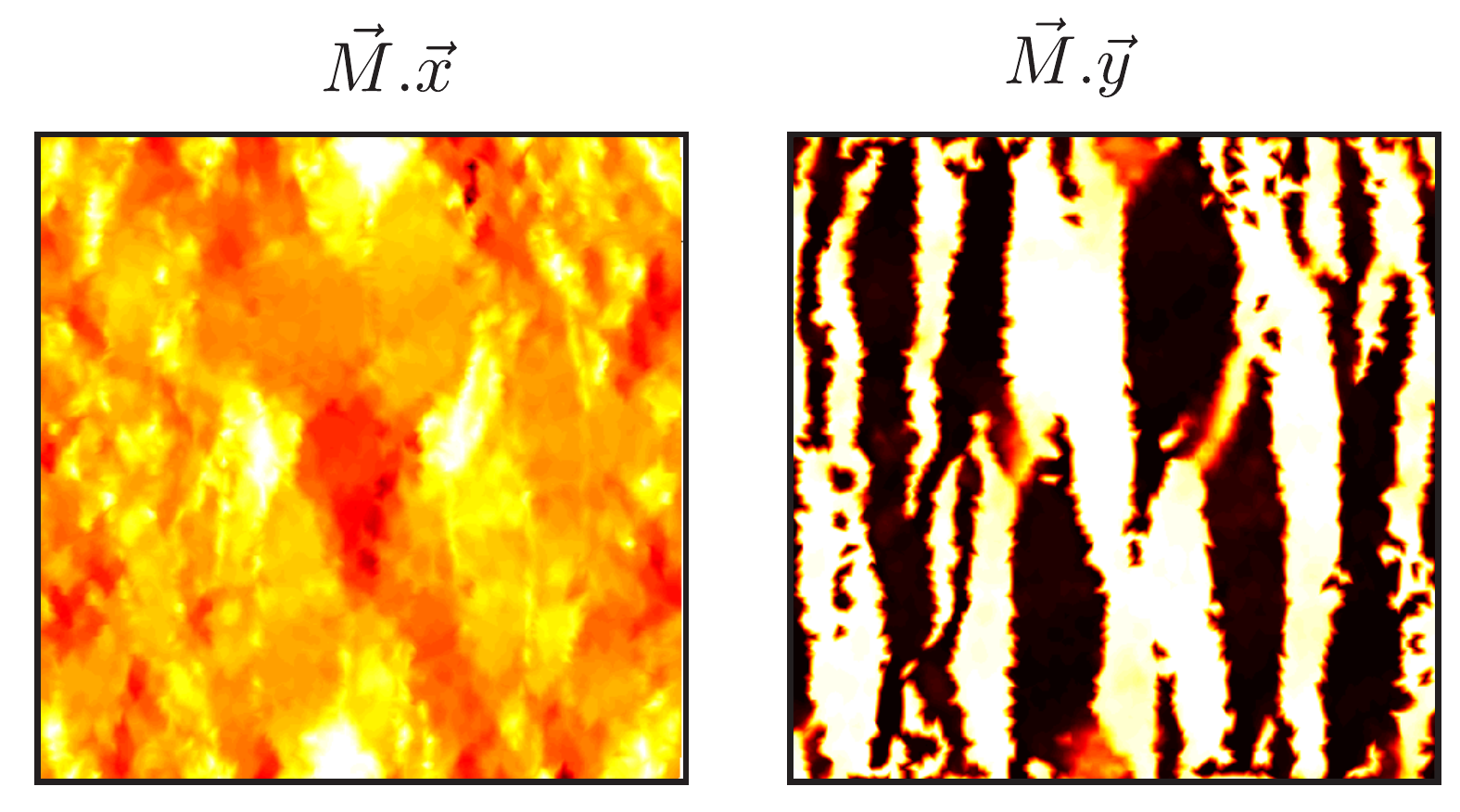}}\qquad
  \subfigure[\label{mSyy} $\Sigma_{yy}=-4,37\ Gpa\quad \Sigma_{xx}=\Sigma_{xy}=0$]{\includegraphics[width=7cm]{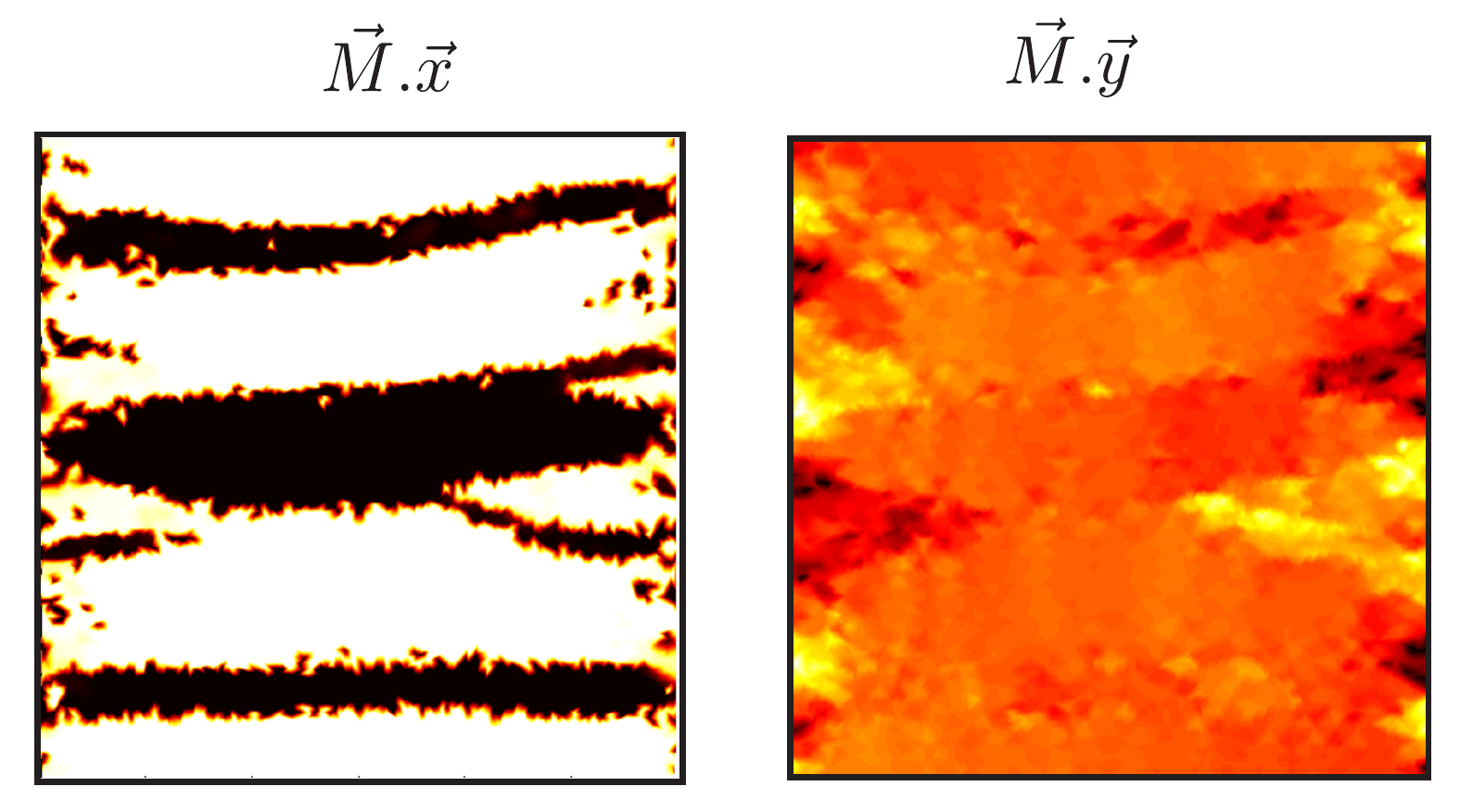}}
%  \subfigure[\label{stress_plan}Contraintes: $\sigma_{max}=13.8Gpa$]{\includegraphics[width=4.5cm]{images_publi/plan_charge}}\quad
% \subfigure[\label{chi}$\chi(\boldsymbol{\sigma})$ avec: $\chi_0=2.2$]{\includegraphics[width=6cm]{images_publi/chi}}  
\caption{ Evolution de la microstructure sous l'influence d'un chargement mcanique. }
\end{figure}

%\begin{equation}
%\chi=\frac{1}{S_{_\Omega}\ \vec{H}_{ext}.\vec{x}}\int_\Omega \vec{M}.\vec{x}\ d_{{\Omega}}
%\end{equation}

Les figures \ref{pSyy} et \ref{mSyy} rendent compte de l'volution de la structure des domaines et des parois magntiques. On peut y  observer qu'en traction (fig. \ref{pSyy}) les domaines qui initialement taient aligns avec le champ ($\vec{M}.\vec{x}$) disparaissent au profit des domaines orthogonaux au champ ($\vec{M}.\vec{y}$) nergtiquement favorables. En compression (fig. \ref{mSyy}) on constate l'effet inverse.
%qui reprsente la carte de susceptibilit obtenue pour chaque tat de contrainte rend compte de la sensibilit de la structure des domaines et des parois magntiques  l'tat de chargement mcanique.  On peut par exemple observer sur l'axe $\boldsymbol{\sigma}_{xx}$, en traction, la forte susceptibilit observe traduit la prfrence des domaines magntiques  s'aligner dans l'axe de traction ici parallle au champ appliqu. Inversement, toujours sur le mme axe $\boldsymbol{\sigma}_{xx}$ la compression provoque une forte chute de la susceptibilit qui traduit une rorganisation de microstructure magntique en faveur des domaines orthogonaux champ impos.

\section{Conclusion}

La mise en \oe uvre du couplage magnto-lastique dans le cadre d'un calcul micromagntique a montr son efficacit.  Sur un monocristal nous avons montr que l'anisotropie induite par l'tat de contrainte intrinsque stabilise la microstructure magntique, en particulier la taille des parois magntiques. La nouvelle libert d'explorer diffrents tats de chargement dans le plan de contrainte (traction/compression uniaxial), nous a permis de mettre en vidence l'volution du comportement magntique, et la sensibilit de la microstructure magntique  l'tat de contrainte.  Dans l'optique de la description du comportement magnto-mcanique des aciers DP, la suite de ce travail consistera  introduire des microstructures de plus en plus complexes (fig. \ref{dp}), l'objectif tant  terme d'obtenir des lois d'volution du comportement magntique en fonction de  l'htrognit (fraction d'\^ilots martensitiques, morphologie des \^ilots ).

%%\\ article

% ---------------------------------------------------------------------
\end{document}